\documentclass[english]{IEEEtran}
\usepackage[T1]{fontenc}
\usepackage[latin9]{inputenc}
\usepackage{color}
\usepackage{amstext}
\usepackage{graphicx}

\makeatletter

\providecommand{\tabularnewline}{\\}

\usepackage{steinmetz}
\usepackage{hyperref}

\makeatother

\usepackage{babel}
\begin{document}

\title{Investigating early warning signs of oscillatory instability in simulated
phasor measurements}

\author{\authorblockN{Goodarz Ghanavati, Paul D.~H.~Hines and Taras I.~Lakoba}
\authorblockA{\\
College of Engineering and \\
Mathematical Sciences\\
 University of Vermont\\
 Burlington, VT\\
} }
\maketitle
\begin{abstract}
This paper shows that the variance of load bus voltage magnitude in
a small power system test case increases monotonically as the system
approaches a Hopf bifurcation. This property can potentially be used
as a method for monitoring oscillatory stability in power grid using
high-resolution phasor measurements. Increasing variance in data from
a dynamical system is a common sign of a phenomenon known as critical
slowing down (CSD). CSD is slower recovery of dynamical systems from
perturbations as they approach critical transitions. Earlier work
has focused on studying CSD in systems approaching voltage collapse;
In this paper, we investigate its occurrence as a power system approaches
a Hopf bifurcation.

\end{abstract}

\section{Introduction}

After several recent blackouts in North America \cite{Southwest,August2003}
and Europe \cite{Bacher:2003}, there is increasing motivation to
use phasor measurement unit (PMU) data for improving reliability in
the electricity industry. For example, the September 8, 2011 Southwestern
US blackout  report \cite{Southwest} recommended the use of PMU data
to increase situational awareness. It emphasized that PMUs may prove
increasingly important in identifying and monitoring for signs of
grid stress, such as dangerous oscillations.

One of the most important grid conditions that operators need to monitor
for is oscillatory stability. Oscillatory stability problems are typically
associated with a pair of complex eigenvalues crossing the imaginary
axis of the complex plane after a system undergoes a contingency \cite{kundur1994power}.
Oscillatory stability problems in a power system can be either local
or global. Local plant mode involves rotor angle oscillations of a
generator against the rest of the system. Global problems known as
inter-area mode oscillations involve generator  in one area swinging
against generators in another area \cite{kundur2004definition}. Inter-area
oscillations can have widespread effects. Incidents of undamped inter-area
oscillations have occurred in many different power grids such as August
10, 1996 blackout in Western North America \cite{Kosterev:1996} and
May 1, 2005 in southeastern Europe \cite{mantzarisanalysis}. There
has been substantial research on fundamentals of this phenomenon in
literature such as \cite{klein1991fundamental,ajjarapu1992bifurcation}.
Reference \cite{klein1991fundamental} analyzes the effects of excitation
systems, loads and DC links on inter-area oscillations. Reference
\cite{ajjarapu1992bifurcation} reveals the existence of stable and
unstable periodic solutions in power system models using the bifurcation
theory.

Numerous methods are proposed for monitoring inter-area oscillations
such as \cite{browne2008comparative}\nocite{canizares2004linear}--\cite{kakimoto2006monitoring}.
Reference \cite{browne2008comparative} compares the Prony analysis
and Hilbert transform methods for modal identification. In \cite{canizares2004linear},
a linear index is developed for identifying Hopf bifurcations, based
on eigenvalues and singular values of state matrix. Reference \cite{kakimoto2006monitoring}
estimates frequency and damping of the inter-area mode from the Fourier
spectrum of phasor measurements. These, while valuable, have limitations
with regard to measurement noise, calculation speed, accuracy and
discriminating between similar modes \cite{browne2008comparative}.
In this paper, we show that changes of statistics of some of system
variables can be a potentially helpful complement to existing methods.

As dynamical systems approach a critical transition, they  increasingly
recover more slowly from perturbations. This phenomenon, known as
critical slowing down (CSD), is primarily caused by  reduced damping
as a system's eigenvalues approach the right-half plane. Increasing
variance and autocorrelation are two common signs of CSD phenomenon
in dynamical systems \cite{scheffer2009early}. The increase in these
two statistics have been observed in many different dynamical systems
\cite{scheffer2009early}, and are frequently suggested as early warning
signs of critical transitions \cite{kuehn2011mathematical}. However,
recent research has shown that CSD signs are not universal \cite{boerlijst2013catastrophic,hastings2010regime};
they may not be observable in all system variables and under all conditions.
Therefore, it is important to carefully choose which variable(s) to
monitor, in order to effectively use autocorrelation or variance as
early warning signs of bifurcation.

Prior research \cite{cotilla2012predicting}\nocite{podolsky2012random,ghanavati2013calculation,podolsky2013critical}--\cite{ghanavati2013understanding}
has shown that the signs of CSD occur in power systems, in the vicinity
of saddle-node bifurcations. In \cite{ghanavati2013understanding},
the present authors showed that CSD signs do not appear in all variables
in the vicinity of a saddle-node bifurcation, in several power system
models. 

In this paper, we investigate changes in autocorrelation and variance
of system variables as a power system model approaches a Hopf bifurcation.
The results show that increasing variance of bus voltage magnitudes
is a good early warning sign of Hopf bifurcation. Thus, monitoring
for this value could potentially be used as an indicator of oscillatory
stability problems in power system. Sec.~\ref{sec:Simulation-and-results}
presents the simulation and results of study of changes in autocorrelation
and variance of a Three-bus test case in the vicinity of Hopf bifurcation.
Sec.~\ref{sec:Discussion} is a discussion on the results presented
in Sec.~\ref{sec:Simulation-and-results}. Sec.~\ref{sec:Conclusion}
highlights the results and contributions made in this paper.

\renewcommand\[{\begin{equation}}

\section{Simulation and results\label{sec:Simulation-and-results}}

This section presents the simulation of a small power system test
case, with which we study the occurrence of CSD  in the vicinity of
a Hopf bifurcation. First, we present the test case and the simulation
method used. Then, the changes of variances and autocorrelations of
system variables in the vicinity of Hopf bifurcation are shown.

\subsection{Test Case and Simulation}

Fig.~\ref{3-bus-test-system} shows the single-line diagram of a
Three-bus test system model \cite{ghasemi2006line} under study. The
two generators in this system are modeled with a standard sixth order
generator model \cite{kundur1994power}, and  are equipped with exciters.
A governor is connected to the first generator. The system data are
given in Appendix A. \textcolor{black}{W}e simulated this system using
the power system analysis toolbox (PSAT) \cite{milano2005open}. Here,
we assume that the load power varies stochastically, with normally
distributed fluctuations. However, since the variance of white noise
is infinite, we assumed that the load perturbations have finite correlation
time. We also assumed that the correlation time of noise is negligible
relative to the response-time of the system. Numerically, this correlation
time is assumed to be equal to the integration time step of (\ref{eq:DAE-f})
below. Adding noise to the system load adds randomness to the system.
Therefore, a set of stochastic differential-algebraic equations (SDAEs)
describe this system:
\begin{eqnarray}
\underline{\dot{x}} & = & f\left(\underline{x},\underline{y}\right)\label{eq:DAE-f}\\
0 & = & g\left(\underline{x},\underline{y},\eta\right)\label{eq:DAE-g}
\end{eqnarray}
where $\underline{x},\underline{y}$ represent the vector of differential
and algebraic variables respectively, $\eta$ is the gaussian random
variable added to the load $\eta\sim\mathcal{N}\left(0,0.01\right)$,
$f,g$ represent the set of differential and algebraic equations of
the system, respectively. A subset of algebraic equations are power
flow equations, into which the noise is added:
\begin{eqnarray}
P_{k}-P_{k0}\eta & = & V_{k}\cdot\label{eq:PF-P}\\
 &  & {\displaystyle \sum_{m=1}^{n}}{\textstyle \left(G_{km}V_{m}\cos\theta_{km}+B_{km}V_{m}\sin\theta_{km}\right)}\nonumber \\
Q_{k}-Q_{k0}\eta & = & V_{k}\cdot\label{eq:PF-Q}\\
 &  & {\displaystyle \sum_{m=1}^{n}}{\textstyle \left(G_{km}V_{m}\sin\theta_{km}-B_{km}V_{m}\cos\theta_{km}\right)}\nonumber 
\end{eqnarray}
where $n=3$; $P_{k}$ and $Q_{k}$ are injected active and reactive
power at each bus; $P_{k0},Q_{k0}$ are constant values; $G_{km}$
and $B_{km}$ are the conductance and the susceptance of the line
between bus $k$ and bus $m$; $V_{m}$ is the voltage magnitude of
bus $m$; $\theta_{km}=\theta_{k}-\theta_{m}$, where $\theta_{k},\theta_{m}$
are voltage angles of buses $k,m$. The differential and algebraic
equations that describe the generator, exciter and turbine governor
are available in \cite{milano2008power}. 

\begin{figure}
\begin{centering}
\includegraphics[width=1\columnwidth]{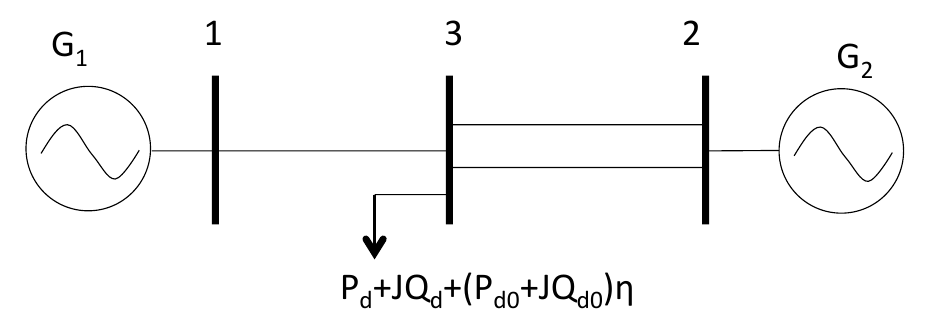}
\par\end{centering}

\caption{\label{3-bus-test-system}Three-bus test system}
\end{figure}

We solved the resulting SDAEs using a fixed-step trapezoidal differential-algebraic
equations solver for different load levels. For each load level, we
simulated the system around the equilibrium. Each load's active and
reactive powers fluctuate around their mean values. The time for each
simulation was 120s and the integration step size was 0.01s. We assumed
that the noise level is constant (i.e. $P_{k0},Q_{k0}$ are constant
in (\ref{eq:PF-P}), (\ref{eq:PF-Q})) when the load is varied, so
as to make sure that the increase of variables' variances is not due
to the increase of the noise level. At the end of the simulation,
we subtracted means of the time-series of the algebraic and differential
variables before calculating their variances and autocorrelations.
For each load level, we ran simulations 100 times, and calculated
the average of variances and autocorrelations of variables. In this
work, we vary $P$ and $Q$ proportionally, so that the power factor
remains constant. 

As the load increases, the system passes through a Hopf bifurcation.
Fig.~\ref{fig:PV-curve-Hopf} shows the PV curve for this system.
Hopf bifurcation occurs before the maximum power transfer limit (the
nose point of the PV curve). Fig.~\ref{fig:Eigenvalues-of-Three-bus}
shows the trajectory of the eigenvalues of the system as the load
increases. Only the three pairs of eigenvalues closest to the right-half
plane are shown. In this and subsequent figures, the dotted line shows
a point close to the bifurcation at which we did eigenvalue analysis
(see Sec.~\ref{sec:Discussion}) to find out why variances and autocorrelations
of the variables show different patterns in the vicinity of the bifurcation.

\begin{figure}
\includegraphics[width=1\columnwidth]{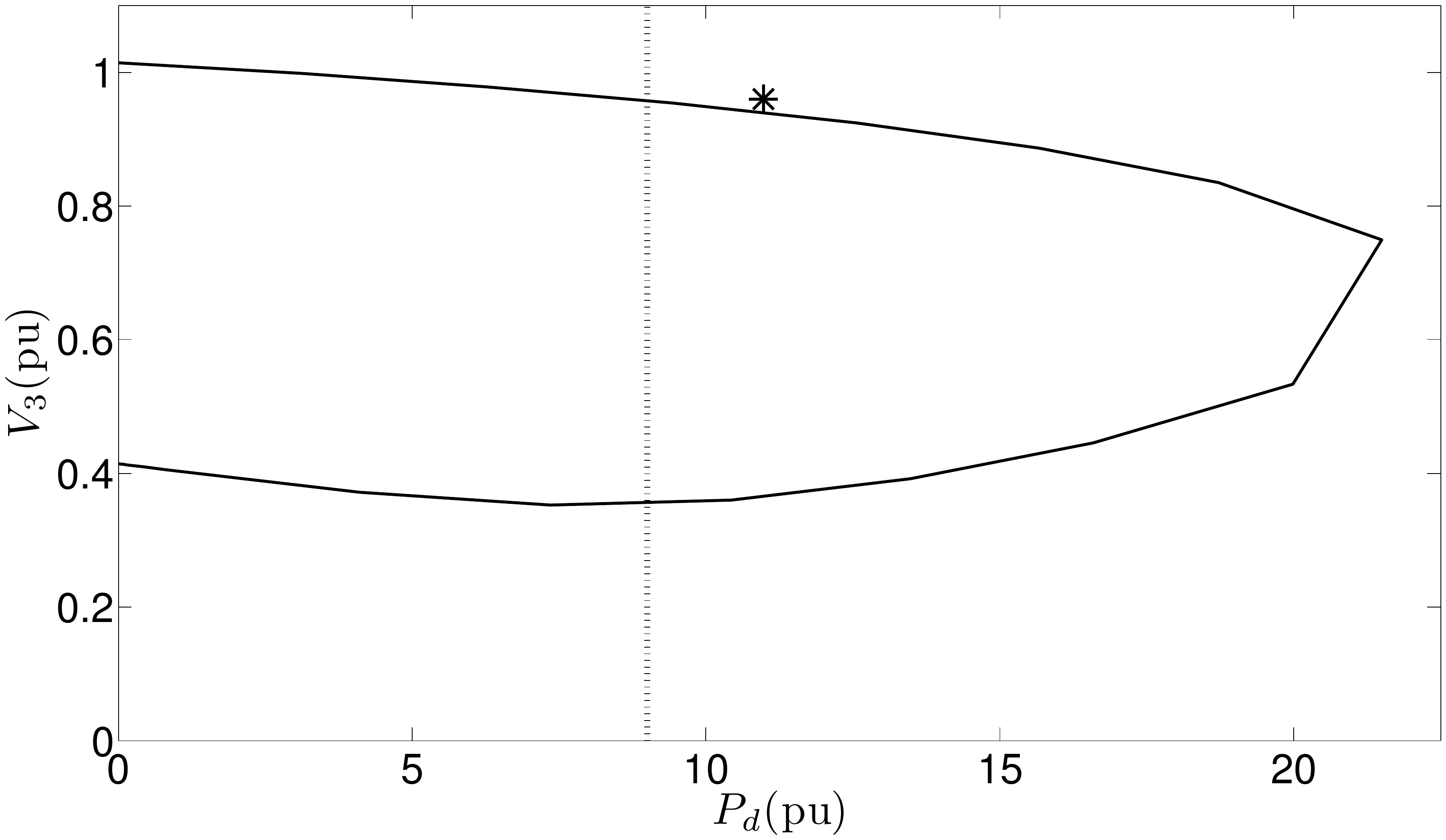}

\caption{\label{fig:PV-curve-Hopf}PV curve for the Three-bus system. {*} denotes
the Hopf bifurcation point. The vertical dotted line shows the nominal
load power $(9\textnormal{{pu}})$.}
\end{figure}

\begin{figure}
\includegraphics[width=1\columnwidth]{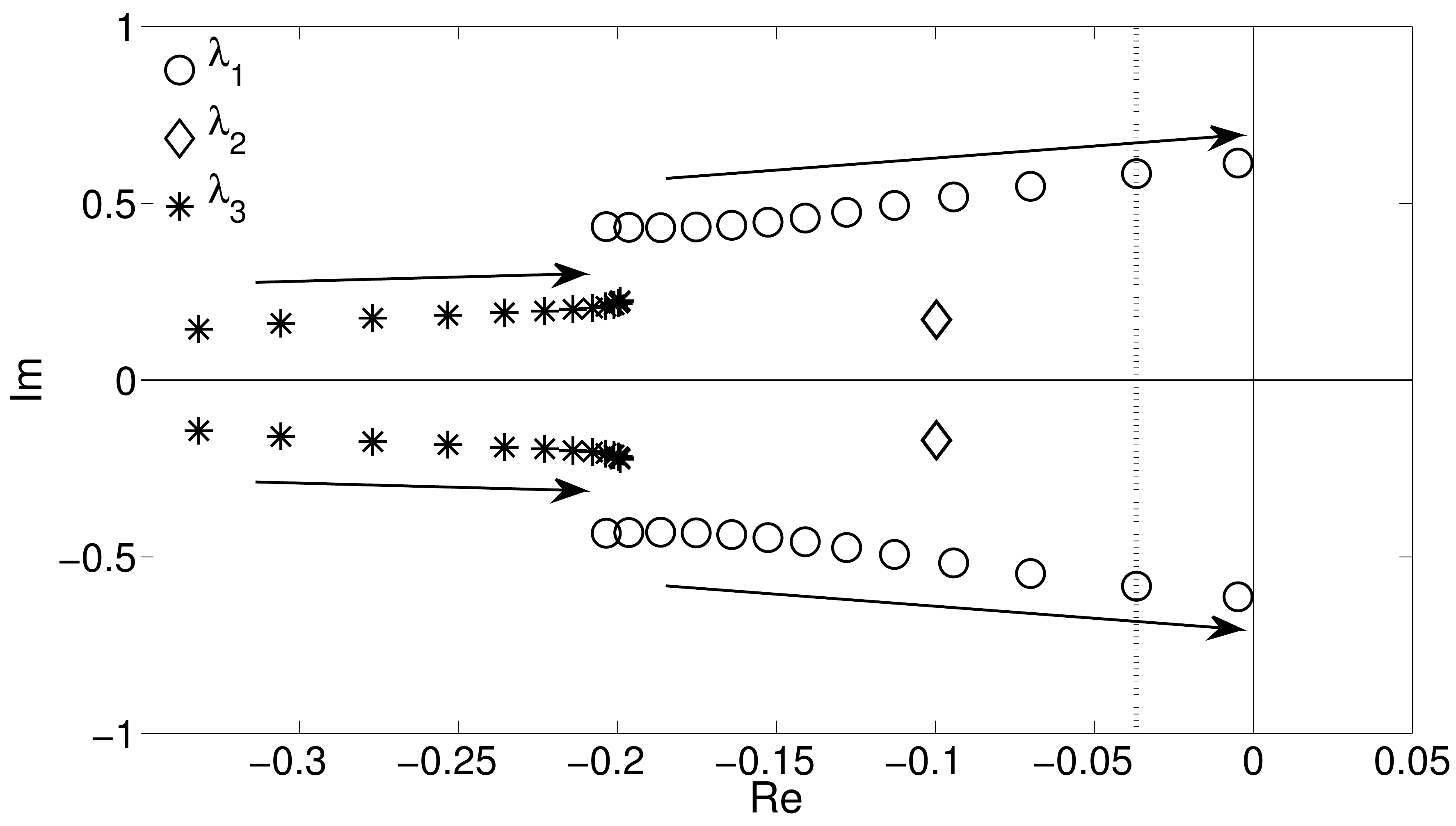}
\vspace{-.2in}
\caption{\label{fig:Eigenvalues-of-Three-bus}Trajectory of the three pairs
of dominant eigenvalues of the Three-bus system as the load is increased.
The arrows show the direction of the eigenvalues' movement in the
complex plane as the load is increased. The increment of bifurcation
parameter $P_{d}$ is $0.9\textnormal{{pu}}$. Near the bifurcation,
the next (fourth) smallest real part of eigenvalues is approximately
$-0.7$.}
\vspace{-.3in}
\end{figure}

\subsection{Autocorrelations and variances of the system variables\label{sub:Autocorrelations-and-variances}}

Figs.~\ref{fig:V}--\ref{fig:omega} show variance and autocorrelation
of several variables that can be measured in real time. Before calculating
the variances and autocorrelations, the variables' means were subtracted
from their values. Horizontal axis is the ratio of $P_{d}$ to the
nominal load ($P_{dn}$).\textcolor{red}{{} }\textcolor{black}{These
figures demonstrate that the bus voltage magnitudes are the only variables
whose variances show a monotonic and, importantly, gradual increase
over the entire range of load values. Among the angle variables, only
the variance of $\theta_{3}$ shows a monotonic and gradual increase
starting with $P_{d}/P_{dn}\approx0.6$, and its increase is less
pronounced than that of $\sigma_{\Delta V}^{2}$. In contrast to the
variances, the autocorrelations of the voltage magnitudes increase
conspicuously only very near the bifurcation, while variances and
autocorrelations of other variables }(Figs.~\ref{fig:theta}--\ref{fig:omega})
are not even monotonic over most of the range of the load level. Therefore,
among all the measurable quantities in the Three-bus system, only
the load bus voltage variance is a reliable early sign of the bifurcation.
\vspace{-.1in}
\begin{figure}
\includegraphics[width=1\columnwidth]{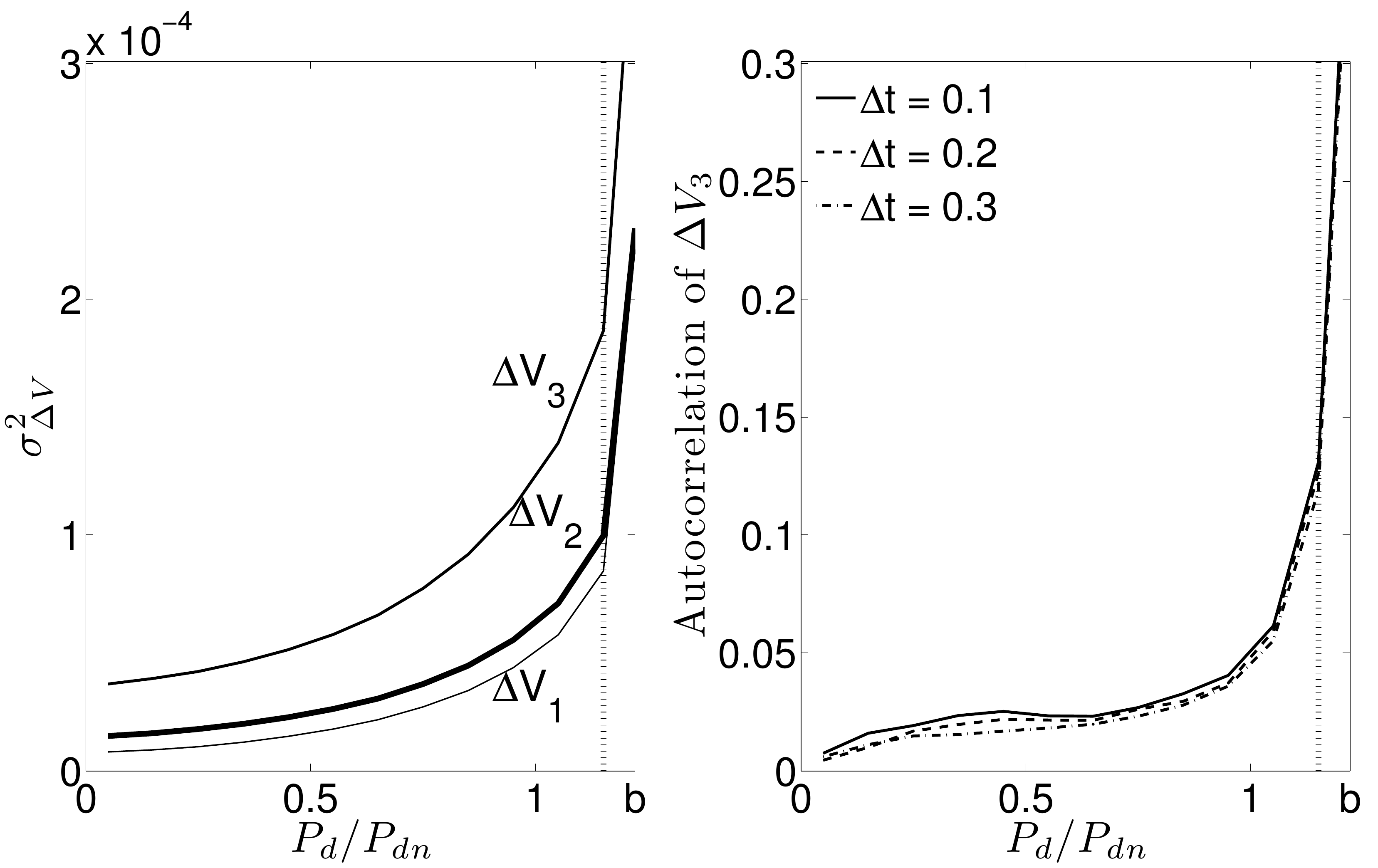}
\vspace{-.2in}
\caption{\label{fig:V}Variance and autocorrelation of the voltage magnitude
of the load bus versus load level. Note that the autocorrelation $\langle\Delta V(t)\Delta V(t+\Delta t)\rangle$
is shown only for $\Delta V_{3}$; it is similar for the other two
voltages.}
\vspace{-.4in}
\end{figure}

\begin{figure}
\vspace{-.2in}
\includegraphics[width=1\columnwidth]{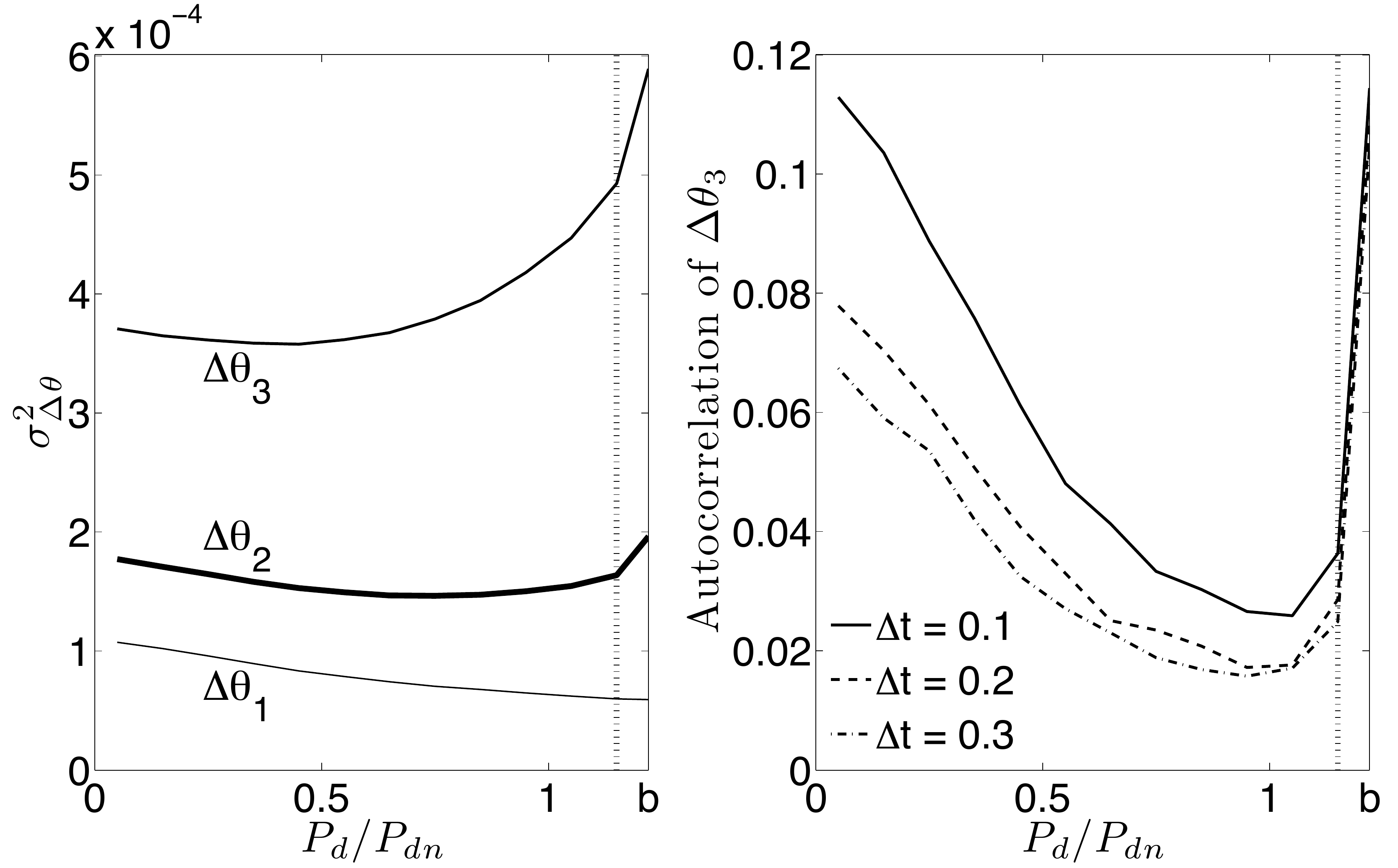}

\caption{\label{fig:theta}Variance and autocorrelation of the voltage angle
of the load bus versus load level. Note that the autocorelation $\langle\Delta\theta(t)\Delta\theta(t+\Delta t)\rangle$
is shown only for $\Delta\theta_{3}$; it is similar for the other
two angles.}

\vspace{-.3in}
\end{figure}

\begin{figure}
\includegraphics[width=1\columnwidth]{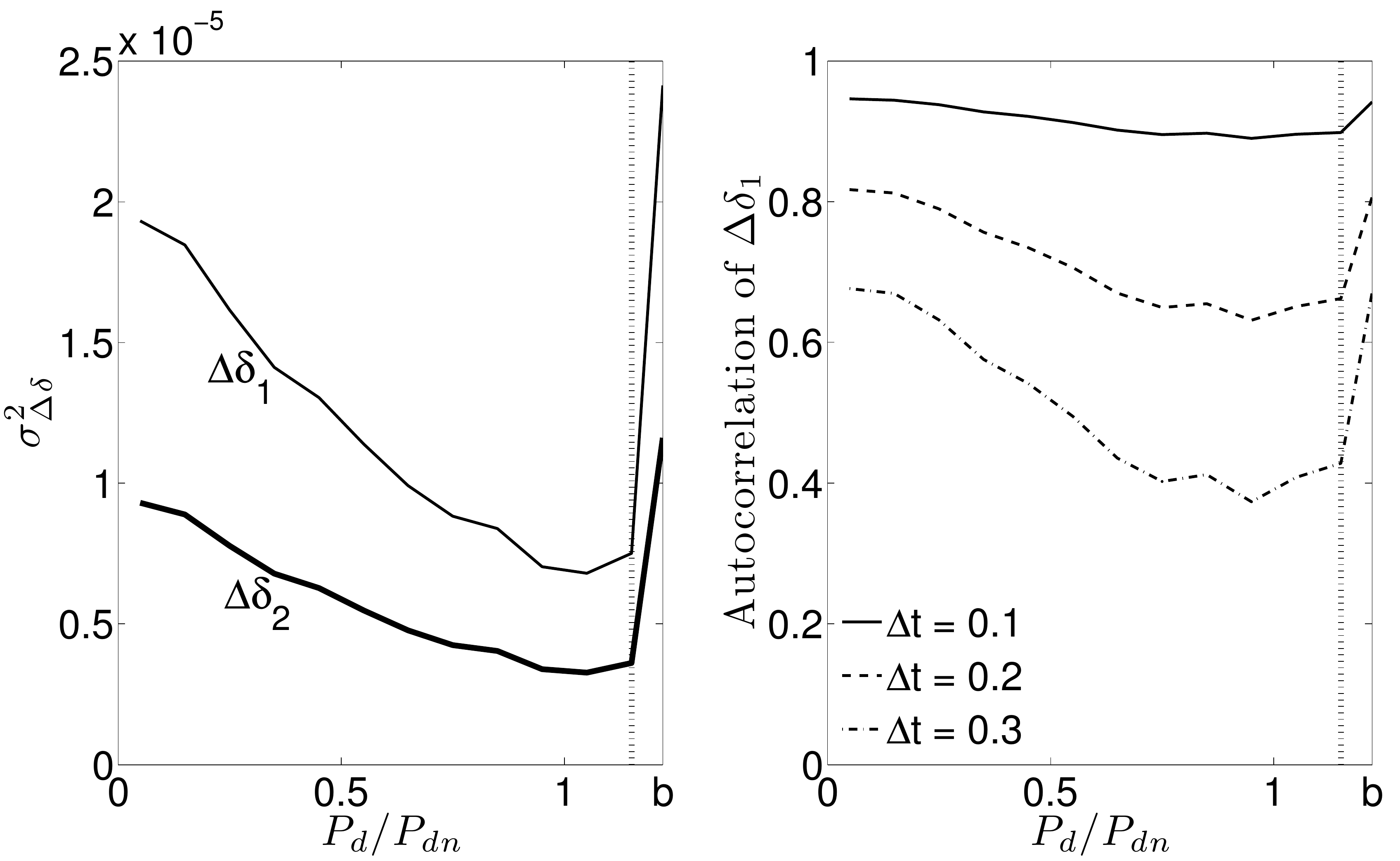}

\caption{\label{fig:delta}Variances and autocorrelations of the generators
rotor angles versus load level. Note that the autocorelation $\langle\Delta\delta(t)\Delta\delta(t+\Delta t)\rangle$
is shown only for $\Delta\delta_{1}$; it is similar for the other
generator angle.}
\vspace{-.1in}
\end{figure}

\begin{figure}
\includegraphics[width=1\columnwidth]{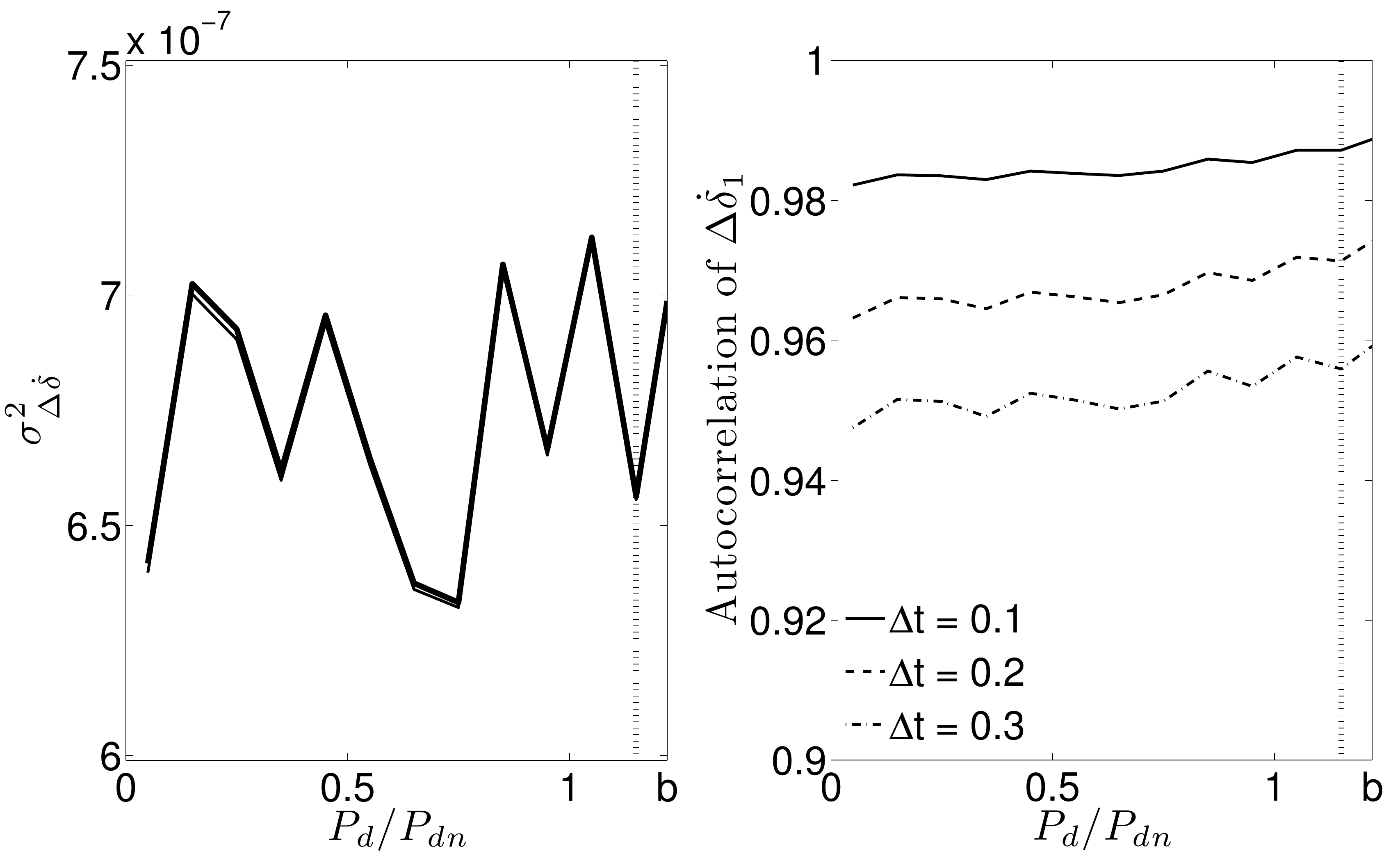}

\caption{\label{fig:omega}Variances and autocorrelations of the generators
speed deviations versus load level. Variances of $\Delta\dot{\delta}_{1}$
and $\Delta\dot{\delta}_{2}$ are very close, so their difference
is not observable in the left-hand side panel. Note that the autocorelation
$\langle\Delta\dot{\delta}(t)\Delta\dot{\delta}(t+\Delta t)\rangle$
is shown only for $\Delta\dot{\delta}_{1}$; it is similar for the
other generator speed.}

\vspace{-.2in}
\end{figure}

\section{Discussion\label{sec:Discussion}}

From the description in Sec.~\ref{sub:Autocorrelations-and-variances},
we conclude that a good early warning sign of a Hopf bifurcation in
power systems is the same as for saddle-node bifurcation \cite{ghanavati2013understanding}.
We emphasize the word ``early''. Indeed, it is well-known that variance
and autocorrelation of most variables increases according to certain
universal laws near a bifurcation \cite{kuehn2011mathematical}. However,
our results demonstrate that only a small subset of such variables
- namely, the bus voltage magnitudes' variance - exhibits a consistent
increase sufficiently far from the bifurcation. Therefore, only these
variables can serve as a useful warning sign, which can potentially
be detected early enough to avert a system collapse. 

Let us point out that this conclusion is supported by the eigenvalue
analysis. It is well-known that right eigenvectors of the state matrix
give the relative activity of state (i.e. differential) variables
when the corresponding mode is excited \cite{kundur1994power}. Eigenvectors
are obtained from linearization of (\ref{eq:DAE-f}), (\ref{eq:DAE-g}),
whereby these equations reduce to:
\begin{equation}
\Delta\dot{\underline{x}}=A\Delta\underline{x}\label{eq:state eq}
\end{equation}
where $A$ is:
\begin{equation}
A=f_{x}-f_{y}g_{y}^{-1}g_{x}
\end{equation}
where $f_{x},f_{y},g_{x},g_{y}$ are matrices of partial derivatives
of (\ref{eq:DAE-f}) and (\ref{eq:DAE-g}) with respect to the differential
and algebraic variables. The solution of (\ref{eq:state eq}) is represented
as:
\begin{equation}
\Delta\underline{x}=\phi\underline{z}\label{eq:dxz}
\end{equation}
where $\phi$ is a matrix whose columns are right eigenvectors of
the state matrix, $\underline{z}$ is the time-dependent vector of
transformed state variables such that each variable is associated
with only one mode \cite{kundur1994power}. In order to determine
the relative activity of algebraic variables, we linearized (\ref{eq:DAE-g})
(with no noise in the load):
\begin{equation}
\Delta\underline{y}=-g_{y}^{-1}g_{x}\Delta\underline{x}\label{eq:dydx}
\end{equation}
Then (\ref{eq:dxz}) and (\ref{eq:dydx}) yield:
\begin{equation}
\Delta\underline{y}=C\underline{z}\label{eq:dyz}
\end{equation}
where $C=-g_{y}^{-1}g_{x}\phi$. The columns of $C$ give the relative
activity of algebraic variables in corresponding modes. 

To identify the most ``active'' variables, one looks for a group
of the entries of the eigenvectors (columns of $\phi)$ for the state
variables, or the columns of matrix $C$ for the algebraic variables.
Strictly speaking, one should do so for all the dominant modes, whose
eigenvalues have the smallest real part, because their response to
an external disturbance (e.g. noise in the load) would decay most
slowly. However, out of the three dominant modes shown in Fig.~\ref{fig:Eigenvalues-of-Three-bus}
we have focused only on the one with the smallest real part for the
specific value of load $P_{d}=10.4\textnormal{{pu}}$ (see the vertical
dotted line in Fig.~\ref{fig:Eigenvalues-of-Three-bus}), and demonstrate
that even such restricted analysis agrees with the results provided
by Figs.~\ref{fig:V}--\ref{fig:omega}. In Table~\ref{tab:Relative-activity-of}
we show the magnitudes of only those entries of the corresponding
column of $\phi$ and of $C$ which can be directly measured. There
are other entries as well, which explains why the displayed entries
do not satisfy the conventional normalization:
\begin{equation}
\left|u_{1}\right|^{2}+\left|u_{2}\right|^{2}+....=1\label{eq:normalize}
\end{equation}
where $u_{i}$ is the i-th entry of a column of $\phi$ or $C$. We
see that the ``activity'' of the state variables $\delta_{1,2},\dot{{\delta_{1,2}}}$
is too small compared to that of other state variables. This is reflected
in Figs.~\ref{fig:delta} and \ref{fig:omega} by the fact that these
variables do not show any substantial increase in variance except
perhaps very near the bifurcation point. The ``activity'' of voltage
magnitudes $V_{1,2,3}$ and angles $\theta_{1,2,3}$ is also not very
high (in light of the normalization (\ref{eq:normalize})). Yet, the
``activity'' of $V_{1,2,3}$ relative to other variables is, apparently,
high enough for the variance of $V_{1,2,3}$ to exhibit conspicuous
increase near the bifurcation. Note that of the three angles, $\theta_{3}$
has the largest ``activity'', and its variance's increase is comparable
to that of $V_{1,2,3}$. The other two angles have too small ``activities'',
and their variances do not show monotonic growth as $P_{d}$ approaches
the Hopf bifurcation value.

\begin{table}
\caption{\label{tab:Relative-activity-of}Relative activity of differential
and algebraic variables in dominant mode }

\centering{}%
\begin{tabular}{cccccc}
\hline 
$\theta_{1}$ & $\theta_{2}$ & $\theta_{3}$ & $V_{1}$ & $V_{2}$ & $V_{3}$\tabularnewline
\hline 
0.0081 & 0.0631 & 0.0946 & 0.1525 & 0.1476 & 0.1664\tabularnewline
$\delta_{1}$ & $\delta_{2}$ & $\dot{{\delta_{1}}}$ & $\dot{{\delta_{2}}}$ &  & \tabularnewline
\cline{1-4} 
$0.0661$ & $0.0254$ & $1e-4$ & $4e-5$ &  & \tabularnewline
\end{tabular}
\vspace{-.1in}
\end{table}

\section{Conclusion\label{sec:Conclusion}}

In this paper, we showed that critical slowing down occurs in power
system for Hopf bifurcation. As previously shown for the saddle-node
bifurcation, the results show that CSD signs are better observable
in some variables than others. We showed that this occurs because
fluctuations of some variables are more aligned with the direction
of dominant mode. Specifically, we found that variance of load bus
voltage magnitude is a good early warning sign of Hopf bifurcation.
This property along with the availability of fast PMU measurements
can potentially help in developing a method for monitoring of oscillatory
stability in power grid using phasor measurements.

\appendices

\section{System data}

System base power is 100 MVA.

\noindent Nominal load and generation:\\
$P_{d}=900\textrm{MW},Q_{d}=300\textrm{MVAR},P_{g2}=400\textrm{MW}$\\

\noindent Synchronous generator:

\noindent \begin{center}
\begin{tabular}{ccccc}
\hline 
Bus no. & Base MVA & $r$(pu) & $X_{d}$(pu) & $X'_{d}$(pu)\tabularnewline
\hline 
1 & 555.5 & 0 & 1.81 & 0.3\tabularnewline
2 & 700 & 0 & 1.81 & 0.3\tabularnewline
\hline 
 & $X"_{d}$(pu) & $T'_{do}$(s) & $T"_{do}$(s) & $X_{q}$(pu)\tabularnewline
\hline 
1 & 0.217 & 7.8 & 0.022 & 1.76\tabularnewline
2 & 0.217 & 7.8 & 0.022 & 1.76\tabularnewline
\hline 
 & $X'_{q}$(pu) & $X"_{q}$(pu) & $T'_{qo}$(s) & $T"_{qo}$(s)\tabularnewline
\hline 
1 & 0.61 & 0.217 & 0.9 & 0.074\tabularnewline
2 & 0.61 & 0.217 & 0.9 & 0.074\tabularnewline
\hline 
 & $M$(s) & $D$(pu) &  & \tabularnewline
\hline 
1 & 9.06 & 0 &  & \tabularnewline
2 & 13.06 & 0 &  & \tabularnewline
\end{tabular}\\

\par\end{center}

\noindent Exciter:\\
Exciter model is PSAT's Type III model \cite{milano2008power}.

\noindent \begin{center}
\begin{tabular}{cccccc}
\hline 
Gen. no. & $v_{max}^{f}$ & $v_{min}^{f}$ & $K_{0}$ & $T_{2}$ & $T_{1}$\tabularnewline
\hline 
1 & 40 & -40 & 20 & 12 & 1\tabularnewline
2 & 40 & -40 & 20 & 12 & 1\tabularnewline
\hline 
 & $v_{0}^{f}$ & $S_{0}$ & $T_{e}$ & $T_{r}$ & \tabularnewline
\hline 
1 & 0 & 0 & 0.04 & 0.05 & \tabularnewline
2 & 0 & 0 & 0.04 & 0.05 & \tabularnewline
\end{tabular}\\

\par\end{center}

\noindent Turbine Governor:\\
Turbine Governor model is PSAT's Type II model \cite{milano2008power}.

\noindent \begin{center}
\begin{tabular}{cccccc}
\hline 
Gen. no. & $R$ & $P_{max}$ & $P_{min}$ & $T_{2}$ & $T_{1}$\tabularnewline
\hline 
1 & 0.2 & 10 & 0.3 & 5 & 0\tabularnewline
\end{tabular}
\par\end{center}

\renewcommand\]{\end{equation}}

\bibliographystyle{IEEEtran}
\bibliography{biblio}

\section*{Author biographies}
\vfill
\begin{IEEEbiographynophoto}{Goodarz Ghanavati} (S`11) received the B.S. and M.S. degrees in Electrical Engineering from Amirkabir University of Technology, Tehran, Iran in 2005 and 2008, respectively. Currently, he is pursuing the Ph.D. degree in Electrical Engineering at University of Vermont. His research interests include power system dynamics, PMU applications and smart grid.
\end{IEEEbiographynophoto}
\vspace{-.4in}

\begin{IEEEbiographynophoto}{Paul D.~H.~Hines} (S`96,M`07) received the Ph.D. in Engineering and Public Policy from Carnegie Mellon University in 2007 and M.S. (2001) and B.S. (1997) degrees in Electrical Engineering from the University of Washington and Seattle Pacific University, respectively.\\
He is currently an Assistant Professor in the School of Engineering, and the Dept. of Computer Science at the University of Vermont, and a member of the adjunct research faculty at the Carnegie Mellon Electricity Industry Center. Formerly he worked at the U.S. National Energy Technology Laboratory, the US Federal Energy Regulatory Commission, Alstom ESCA, and Black and Veatch. He currently serves as the vice-chair of the IEEE Working Group on Understanding, Prediction, Mitigation and Restoration of Cascading Failures, and as an Associate Editor for the IEEE Transactions on Smart Grid. He is National Science Foundation CAREER award winner.
\end{IEEEbiographynophoto}
\vspace{-.4in}

\begin{IEEEbiographynophoto}{Taras I. Lakoba} received the Diploma in physics from Moscow State University, Moscow, Russia, in 1989, and the Ph.D. degree in applied mathematics from Clarkson University, Potsdam, NY, in 1996.\\
In 2000 he joined the Optical Networking Group at Lucent Technologies, where he was engaged in the development of an ultralong-haul terrestrial fiber-optic transmission system. Since 2003 he has been with the Department of Mathematics and Statistics of the University of Vermont. His research interests include multichannel all-optical regeneration, the effect of noise in fiber-optic communication systems, stability of numerical methods for nonlinear wave equations, and perturbation techniques.
\end{IEEEbiographynophoto}
\vspace{-.4in}

\vfill
\end{document}